\documentclass[%
 reprint,
nofootinbib,
 amsmath,amssymb,
 aps,
]{revtex4-2}

\usepackage{graphicx}
\usepackage{dcolumn}
\usepackage{bm}
\usepackage{xcolor}

\newcommand{\bk}{\mathbf{k}}
\newcommand{\bq}{\mathbf{q}}

\newcommand{\bP}{\mathbf{P}}
\newcommand{\bp}{\mathbf{p}}
\newcommand{\bA}{\mathbf{A}}

\newcommand{\phiR}{\phi_R}
\newcommand{\phiRstar}{\phi^*_R}

\newcommand{\be}{\mathbf{e}}
\newcommand{\bs}{\mathbf{s}}
\newcommand{\br}{\mathbf{r}}
\newcommand{\bR}{\mathbf{R}}

\newcommand{\del}{\mathbf{\nabla}}

\newcommand{\abs}[1]{\left\vert #1 \right\vert}

\usepackage[colorlinks = true,
            linkcolor = blue,
            urlcolor  = blue,
            citecolor = blue,
            anchorcolor = blue]{hyperref}

\begin{document}

\preprint{APS/123-QED}

\title{Chiral Cavity Control of the Interlayer Exciton Energy Spectrum}

\author{Jonathan Sanchez-Lopez}
 \thanks{These authors contributed equally to this work}
 \affiliation{Department of Physics and Astronomy,\\
 University of California, Los Angeles, CA 90095, USA}

\author{Ze-Xun Lin$^*$}
\thanks{Corresponding author: \href{mailto:zl659@cam.ac.uk}{zl659@cam.ac.uk}}
\affiliation{Division of Physical Sciences, College of Letters and Science, University of California, Los Angeles, CA 90095, USA}
\author{Di Luo}
\affiliation{Department of Electrical and Computer Engineering, University of California Los Angeles, Los Angeles, CA 90095, USA}%

\author{Prineha Narang}
\thanks{Corresponding author: \href{mailto:prineha@ucla.edu}{prineha@ucla.edu}}  
\affiliation{Division of Physical Sciences, College of Letters and Science, University of California, Los Angeles, CA 90095, USA}%
\affiliation{
 Department of Electrical and Computer Engineering, University of California Los Angeles, Los Angeles, CA 90095, USA
}%

\date{\today}

\begin{abstract}
Heterostructures of two-dimensional materials offer a versatile platform to study light-matter interactions of electron and hole gases. By separating electron and hole layers with an insulator long-lived electron-hole bound states known as interlayer excitons can form. We predict that by placing an interlayer exciton in a time-reversal-symmetry-breaking chiral cavity the energy spectrum of an interlayer exciton can be reordered. As a consequence of this reordering the ground state of the interlayer exciton can be driven from an s-orbital to a p-orbital,  effectively changing the symmetry of the electron-hole pair. We present a phase diagram showing the couplings and separations required for a p-orbital excitonic ground state where we predict that larger interlayer separations require higher cavity couplings. We expect these results to be relevant for angular-momentum-tunable, single photon emission physics.
\end{abstract}

\maketitle

\textit{Introduction—} Cavity-QED is a promising avenue for controlling material properties such as enhancement and breakdown of correlated phenomena\cite{topologyOfBS,towardOnDemControl,VariationalThyofNRQED,CavitySC, cavityExciton,topBreakdown, gateTunablePhases, BardasisPolaritons, dynamicalCondensate}. The heterostructures of two-dimensional materials offer a diverse landscape of correlated phases \cite{e-hFluid, stronglyCorrelated,perfectCoulombDrag}, which, when combined with cavity-QED, results in an interesting and rapidly growing field of study.

Breaking of time-reversal symmetry (TRS) is crucial to observe topological phases and topological order \cite{TopExIns,HaldaneModel,FAQHE, IQHE}. Often, TRS is broken explicitly by
the application of an external magnetic field or spontaneously by the electron-electron interaction. Recently, there has been increasing interest in breaking TRS by engineering chiral cavities which host a single circular polarization of light. Chiral cavity designs have been proposed and implemented \cite{JuniCavity,juniRealized,single-HandedCavity, suarez2024chiral, spin-preservingChiralPC,chiralOptics}, and their effects on two-dimensional materials such as graphene have been a topic of recent exploration \cite{grapheneCavity, tbgCavity, MultilayerGrapheneCavity,hybridLightMatter}. 

\begin{figure}[htp]
    \centering
    \includegraphics[width= \linewidth]{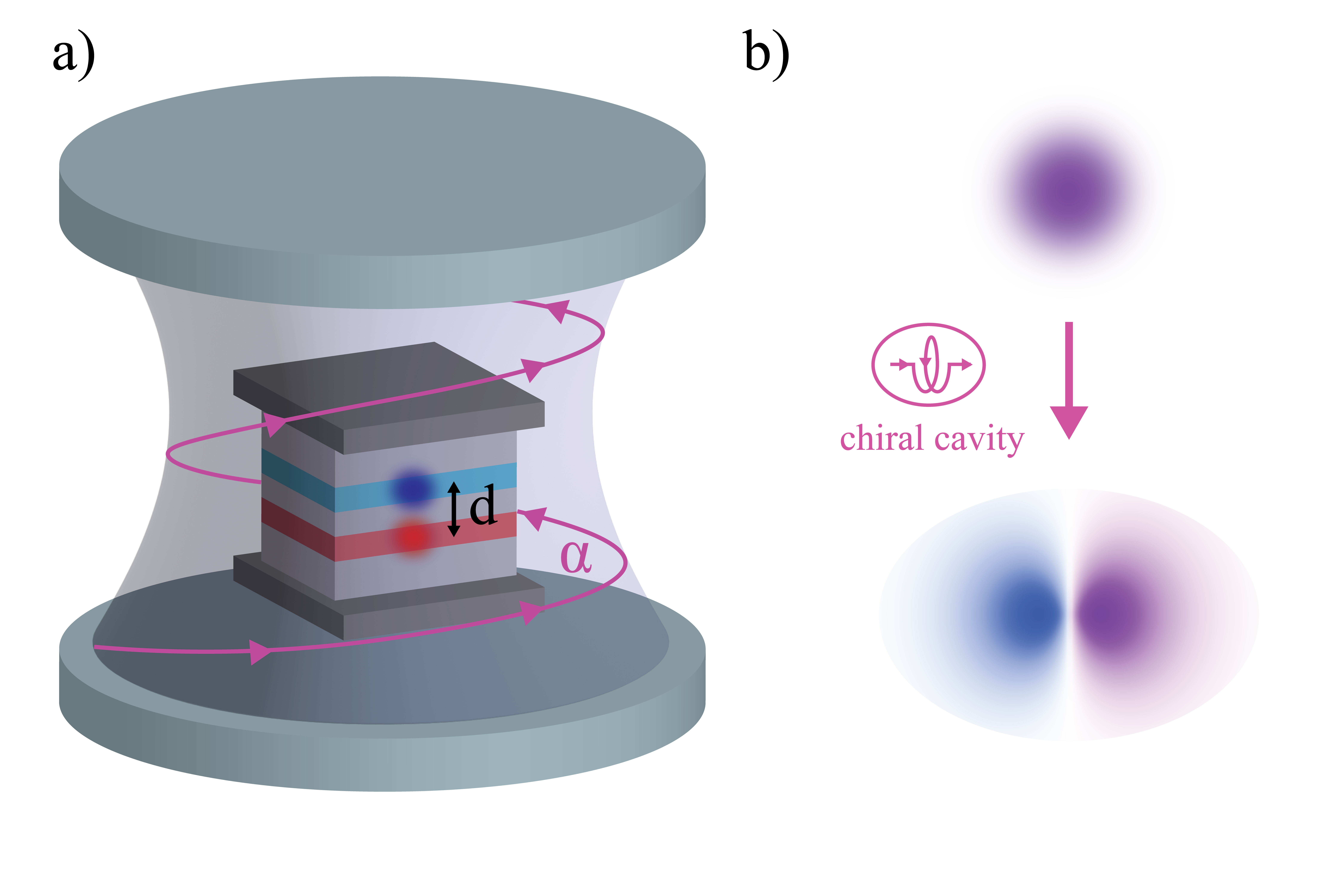}
    \caption{a) Heterostructure in a chiral cavity (gray disks). A device consisting of two H-phase TMD monolayers with interlayer separation d and cavity coupling $\alpha$: an electron (blue) and a hole (red) monolayer separated by insulating hBN (light gray). The TMD bilayer is surrounded by more hBN and encapsuled by two graphene gates (dark gray slabs). In experiment the voltage difference between the gates and TMDs allow for independent control of charge carriers in the TMDs. The cavity hosts purely left- or right-handed photons. b) Chiral cavity coupling can drive the ground state of an exciton from an isotropic s-orbital to an anisotropic p-orbital.} 
    \label{fig: setup}
\end{figure}

In this letter, we discuss the effects of a TRS-breaking chiral cavity on a single interlayer exciton. We have in mind a device like that shown in FIG. \ref{fig: setup}a) where two electrically controlled TMD monolayers with type-II band alignment \cite{emergingExciton} are separated by a hexagonal insulating few-layer Boron Nitride (hBN). Surrounding the TMDs is multilayer hBN; on the top and bottom of this device are graphene monolayers, which serve as gates. Electrical contacts are made between the graphene gates and between the TMDs. The charge carrier densities can be controlled independently in each layer by tuning the voltage difference between the contacts. Experimental realizations of these heterostructure devices have enabled the measurement of exciton binding energy and exciton density through capacitance measurements \cite{stronglyCorrelated}. Further experiments have shown perfect drag of carriers in one layer by driving a current in the other layer \cite{perfectCoulombDrag}.

\textit{Model—}The Hamiltonian for an electron and a hole interacting via Coulomb attraction is given by (in SI units)
\begin{equation}
    H_0 = \frac{p_e^2}{2m_e} + \frac{p_h^2}{2m_h} - \frac{e^2}{4 \pi \epsilon} \frac{1}{\sqrt{(x_e - x_h)^2 + (y_e - y_h)^2 + d^2}}\label{eq: H0}
\end{equation}
Where $p_{e/h}$ are the electron and hole momenta, respectively; $m_{e/h}$ is the effective electron and hole mass, respectively; $x_{e/h}$ are the electron and hole $x$-coordinates, respectively; $y_{e/h}$ is the electron and hole $y$-coordinates, respectively; $\epsilon$ is the electric permittivity of the insulating barrier, and $d$ is the interlayer separation of TMDs. Since TMD have parabolic dispersions \cite{kpTMD} and previous studies have shown that layer twist angles do not change the physics of TMD-insulator-TMD heterostructures \cite{noMoireEffects}, we take this to model the low energy physics of the system.
 The coupling to the cavity's vector potential is given via the replacement $\mathbf{p}\rightarrow \mathbf{p} - q\mathbf{A}$ where $\mathbf{A}$ is the cavity's vector potential given by 
$\mathbf{A} = \sqrt{\frac{\hbar}{\epsilon_C \mathcal{V} 2 \omega _c}}\left(\mathbf{e}_R b_R^\dagger + \mathbf{e}_L b_R\right)$ \cite{JuniCavity} where $\epsilon_C$ is the cavity dielectric constant, $\mathcal{V}$ is the effective
mode volume, $\omega_c$ is the mode frequency, $b_R^\dagger$ is the right-handed photon creation operator, and $\be_R$ and $\be_L$ are the right- and left-handed circularly polarized field vectors —for conventions used refer to the Supplemental Information (SI)—. The precise details of the vector potential's amplitude depend on the cavity design. We assume that the cavity supports a single mode and that the field is homogeneous throughout our device.

We obtain an effective Hamiltonian, $H_\textrm{eff}$,  by tracing photonic field configurations: The partition function of our system is given by a path integral  
over photonic coherent states and one-body position 
and momentum eigenstates for the electron and hole $Z \sim \int \mathcal{D}(\phi^*_R, \phi_R)\exp{(-\frac{S[\phi, \br_e, \bp_e,  \br_h, \bp_h]}{\hbar}})$. Here $S[\phi, \br_e, \bp_e,  \br_h, \bp_h]$ is the action due to the minimally coupled Hamiltonian.
We can perform the photonic integration since 
$S[\phi, \br_e, \bp_e,  \br_h, \bp_h]$ is quadratic in $\phi$. The effective Hamiltonian can be obtained from the resulting expression for the partition function — refer to SI for further details—. The effective Hamiltonian obtained is
\begin{eqnarray}
    &&H_{\mathrm{eff}} =\frac{g^2}{\mu} + \frac{1}{2} \hbar \omega_c +\frac{\bP^2}{2 (m_e + m_h)}  + \frac{\bq^2}{2\Sigma}  - \frac{e^2}{4 \pi \epsilon}\frac{1}{\sqrt{\bs^2 + d^2}}\nonumber\\
    &&- \frac{e^2}{4\pi\epsilon}\left(\frac{g}{2\Sigma \omega_c}\right)^2\left[ \frac{\bs^2-2d^2}{(\bs^2 + d^2)^{5/2}}-\frac{2}{(\bs^2 + d^2)^{3/2}}\frac{\mathbf{L}}{\hbar}\cdot \mathbf{\hat z}\right]\label{eq: bare Heff}
\end{eqnarray}
Here, $\mu$ is the reduced mass $\mu^{-1} = m_e^{-1} + m_h^{-1}$, we call $\Sigma = \mu + g^2/\hbar \omega_c$ the effective reduced mass
, which is the modified reduced mass due to the cavity. $g$, with units of momentum, is given by 
$g^2 = \frac{\hbar e^2}{2\mathcal{V}\omega_c \epsilon_C}.$
$\bP$ is the center of mass momentum in the plane, and $\bq$ and $\bs$ are
the relative in-plane momentum and in-plane separation, respectively.
Finally, $\mathbf{L} = \bs\times \bq$, the relative angular momentum. A detailed derivation is given in the SI.

We are interested in the relative dynamics of motion. Therefore, we focus on 
\begin{eqnarray}
    &&H_{\mathrm{eff}} =\frac{\bq^2}{2\Sigma}  - \frac{e^2}{4 \pi \epsilon}\frac{1}{\sqrt{\bs^2 + d^2}} - \frac{e^2}{4\pi\epsilon}\left(\frac{g}{2\Sigma \omega_c}\right)^2\nonumber\\
    &&\hspace{1cm}\times\left[ \frac{\bs^2-2d^2}{(\bs^2 + d^2)^{5/2}}-\frac{2}{(\bs^2 + d^2)^{3/2}}\frac{\mathbf{L}}{\hbar}\cdot \mathbf{\hat z}\right]\label{eq: H_eff}    
\end{eqnarray}
We note that in the full Hamiltonian, TRS breaking occurs since only right-handed photons are present: Upon time reversal, $\mathcal{T}$, the photonic operators become $\mathcal{T}(b_{R,L}) = -b_{L,R}$. Since $\mathbf{L} \mapsto-\mathbf{L}$ under time-reversal, our $H_\mathrm{eff}$ also breaks TRS. The effective Hamiltonian for left-circularly polarized photons can be found by time-reversing our obtained $H_{\mathrm{eff}}$.

The cavity mediates an attractive (repulsive) interaction when the interlayer separation $d$ satisfies $s>d\sqrt{2}$ ($s<d\sqrt{2}$) as seen in the first term in square brackets. For nonzero angular momentum, states with negative (positive) angular momentum decrease (increase) in energy compared to zero angular momentum states, as seen by the second term in square brackets. Due to the nature of the angular-dependent term, we expect certain bound states with positive angular momentum to become unbound due to their positive contribution to their energy. 

\begin{figure*}
    \centering
    \includegraphics[width=1\linewidth]{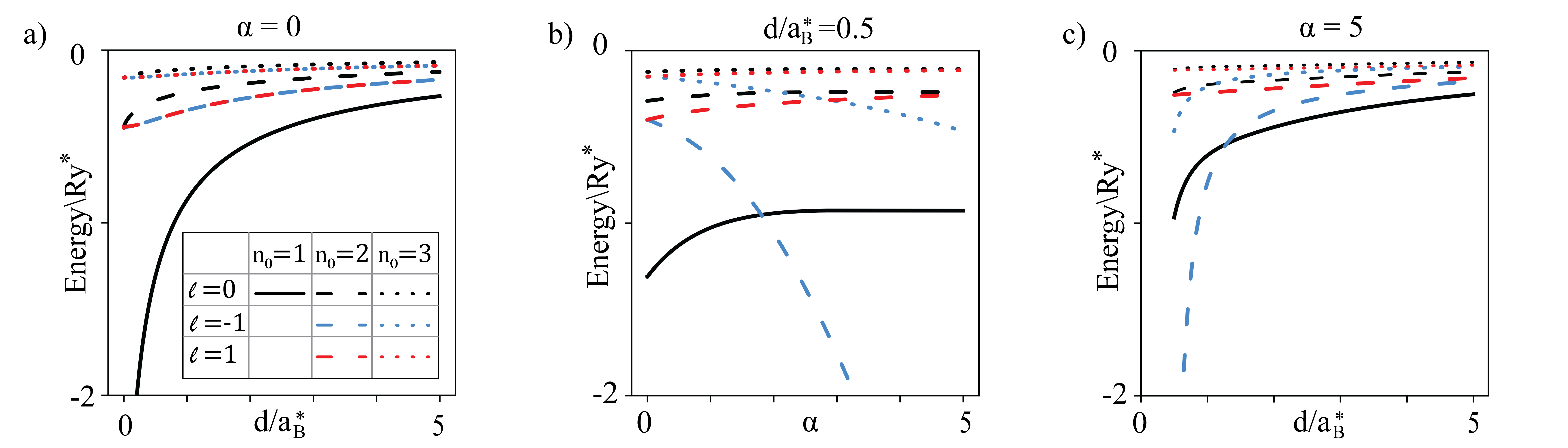}
    \caption{a) First few energy levels of the indirect exciton in the absence of a cavity. The interlayer separation breaks the degeneracy between states with and without angular momentum. Angular momentum states and their time-reversed version remain degenerate. The table in a) is the legend used in a), b) and c). b) and c): effects of variation of $\alpha$ at $d = 0.5 \; a_B^*$ and interlayer separation $d$ at $\alpha = 5$ on the same same levels as a), respectively. The effects of decreasing $\alpha$ are similar to the effects of increasing $d$.}
    \label{fig: joinedFigs}
\end{figure*}

\textit{Analysis—}We preface our analysis by stating that the results found involving the cavity cannot be extrapolated to $d\rightarrow 0$. As interlayer separation decreases, we encounter ground states that diverge in energy, and it is critical that the layers be separated. However, this is not too harsh of a constraint since the electron and hole should be located in different layers to not combine, i.e. there should be at least one layer of insulating material. In the case of hBN, the thickness of a single layer is approximately $0.4$ nm \cite{hBNThickness}. The findings of our work cover such a separation and can therefore be applied to hBN.

As in atomic physics, we present energies in effective Rydbergs and lengths in effective
Bohr radii defined by $1Ry^* = \frac{\Sigma e^4}{2(4\pi \epsilon \hbar)^2}$ and $1a^*_B = \frac{4\pi \epsilon\hbar^2}{\Sigma e^2}$, respectively. We note that in these definitions we use the modified reduced mass $\Sigma$ instead of $\mu = \frac{m_e^* m_h^*}{m_e^*+ m_h^*}$. To provide some sense of scale, let $m = 511 \text{keV}/c^2$ be the bare electron mass, setting $m_e^* = m_h^* = 0.7m$,  $\epsilon = 5\epsilon_0$ and $g=0$ results in $1 Ry^* = 0.19$ eV and $1a_B^* = 0.8$ nm. The dimensionless quantity
\begin{equation}
    \alpha = \frac{1}{2}\left(\frac{g}{\Sigma \omega_c a^*_B}\right)^2 = \frac{\pi e^6}{(4\pi \epsilon_C)(4\pi\epsilon)^2 \mathcal{V}(\hbar \omega_c)^3},
\end{equation}
controls the effects of the cavity on our system, we refer to it as the cavity coupling. How this expression is obtained can be seen in the SI. When there is no cavity, $\alpha = 0$. These definitions effectively reduce the number of tunable parameters to only two: cavity coupling and interlayer separation.

As a final remark, we will denote energy levels by their principal quantum number $n_0$ defined in the limit of no cavity and no interlayer separation ($\alpha\rightarrow0, d \rightarrow0$) as in \cite{2dHydrogen}, for a given angular momentum quantum number $\ell$, $E_{n_0,\ell} <E_{n_0+1,\ell}<0$. For values $d\not = 0, \alpha\not = 0$, $n_0$ is just a label to keep track of various energy levels.

Calculations of the binding energy of interlayer excitons have previously been performed \cite{excitonBinding} as a function of interlayer separation, although in the absence of a cavity. FIG. \ref{fig: joinedFigs}a) shows the the first few energy levels of the exciton as a function of interlayer separation when the cavity is absent. Our binding energies, the $n_0=1, \ell = 0$ energies, agree with such previous studies and expand upon these by including higher energy orbitals. We see that eigenstates with zero angular momentum quantum number, $\ell$, become non-degenerate at $d\not = 0$ —the $\ell=0$ and $\ell\not = 0$ orbitals have different energies— Orbitals with zero angular momentum and principal quantum number $n_0$ have higher energies than their $\ell\not = 0$ counterparts of the same principal quantum number $n_0$. By varying the interlayer separation one can tune the energy difference between exciton energy levels; thus, a spatially indirect exciton provides a playground for tunable energy spectra.

The TRS breaking from the chiral cavity can be observed in FIG. \ref{fig: joinedFigs}b): As we increase $\alpha$ while holding $d$ constant, we see that the $\ell=0,1$ orbitals increase in energy while the $\ell=-1$ states decrease in energy. If we choose a time-reversed cavity where the photons are left-handed, then we see the same trend on the time-reversed orbitals. In other words, we would instead see that the $\ell= 1$ states would decrease in energy while the $\ell= -1,0$ states would increase in energy.

We show in FIG. \ref{fig: joinedFigs}b) the lowest few energy levels of the exciton as a function of $\alpha$ for right-handed photons, the interlayer, $d$, separation is fixed at $d/a_B^* = 0.5 \approx 0.4$ nm —roughly the thickness of an hBN monolayer. As we increase the cavity coupling, the lowest energy state becomes a $\ell= -1$ state at $\alpha \approx 1.8$. As we further increase $\alpha$, the ground state energy continues to decrease while the first excited state's energy increases. We see that the second lowest $\ell= -1$ state also crosses the lowest $\ell= 1$ state at $\alpha \approx 3$. Thus, if we track the angular momenta of the ground, first excited and second excited states as ($\ell_1$, $\ell_2$, $\ell_3$) we follow the path $(0,-1,1)\rightarrow (-1,0,1) \rightarrow(-1,0,-1)$. In principle, one could continue to lower $\ell= -1$ states by increasing $\alpha$, however special care should be taken into account since at higher $\alpha$ the $\ell <-1$ states may become relevant. However, for the range of $\alpha$ considered here, we always report the lowest observed energy levels. Thus, starting from zero coupling, the cavity decreases the difference between the lowest time-reversal symmetric s-orbital and a lowest time-reversal asymmetric p-orbital until the ground state finally becomes the p-orbital.
\begin{figure}[htp]
    \centering
    \includegraphics[width= \linewidth]{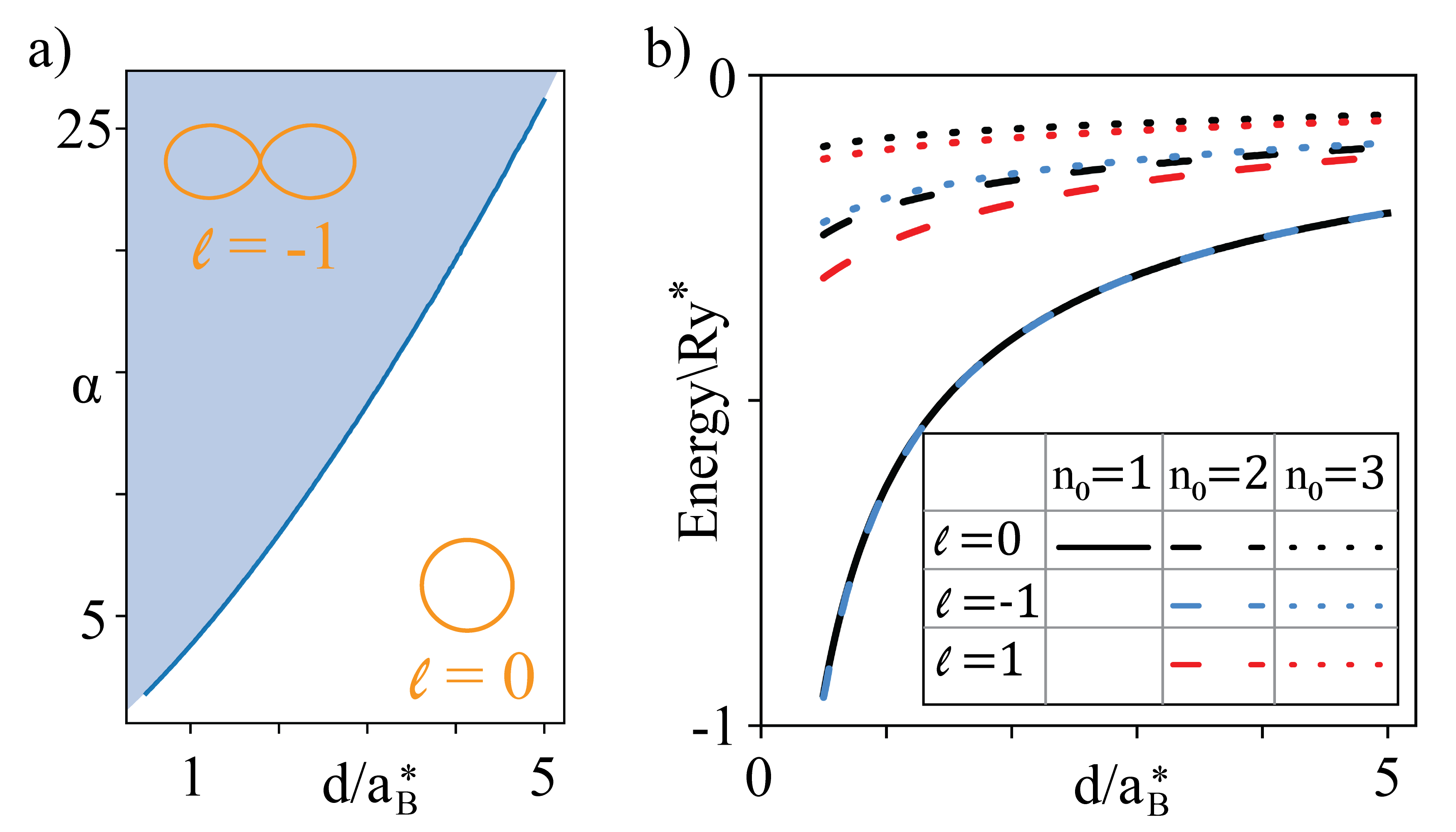}
    \caption{a) Ground State Phase Diagram: as the separation of the layers increases, a higher cavity coupling is needed. The shaded region (blue) indicates parameters where the ground state is a p-orbital. The range of interlayer separations corresponds to typical experimental devices. b) Energies of lowest orbitals along phase boundary from a). Lowest $L = 0$ and lowest $L = -1$ energy levels are degenerate. There are no energy crossings nor are there any degeneracies other than the ground state.} 
    \label{fig: boundaryPlots}
\end{figure}

We now explore the effects of interlayer separation at fixed cavity coupling. In FIG. \ref{fig: joinedFigs}c) we show the effects of interlayer separation on the same energy levels considered thus far when $\alpha = 5$—the highest coupling explored in FIG. \ref{fig: joinedFigs}b)—. As interlayer separation increases, the second excited state transitions from a $\ell = -1$ state to a $\ell = 1$ state at $d \approx 0.7\; a_B^*$. Then, at $d \approx 1.2 \; a_B^*$, the ground state transitions from a $\ell = -1$ state to a $\ell = 0$ state. If we track the angular momentum of the first three levels as done previously, we notice that we have followed the reverse path, i.e., $(-1,0,-1)\rightarrow (-1,0,1) \rightarrow (0,-1,1)$. Therefore, increasing the interlayer separation has an effect similar to reducing $\alpha$, this is expected due to the approximate $d^{n}$-dependence on the cavity-mediated terms with $n\leq-3$. The difference between varying the two parameters is that in the case of varying the interlayer separation, all energy levels increase or decrease. However, in the case of varying $\alpha$ a set of energy levels increases while the other set decreases.

We expect that experimental realizations of our system will not have tunable interlayer separations, rather to achieve varying interlayer separations multiple devices will need to be fabricated. Therefore we explore the $\alpha$-$d$ parameter space to find the conditions for a $\ell = -1$ ground state for various values of $d$. FIG. \ref{fig: boundaryPlots}a) shows the boundary between the two phases, we will denote the boundary between the two phases as $\alpha_c(d)$, i.e. the critical value of $\alpha$ for a given interlayer separation, $d$. We see that as one increases the interlayer separation one must have a larger $\alpha$ to obtain a $\ell=-1$ ground state. This is consistent with the an increase in $d$ behaving similarly to a decrease in $\alpha$. We see in such a phase diagram that in the limit of $\alpha = 0$, we recover the usual s-orbital ground state.

Finally, we explore the energy spectrum along the phase boundary. In FIG. \ref{fig: boundaryPlots}b) We show the energies of the first few energy levels as a function of $\alpha_c(d)$, we see that there are no crossings and verify that the $\ell=0$ and $\ell = -1$ states are degenerate. The energy of all orbitals increases with increasing $\alpha_c(d)$, behavior that is qualitatively the same as that for FIG. \ref{fig: joinedFigs} a) except that all degeneracies have been removed —other than the ground state.

\textit{Discussion and Outlook—}We have argued that the TRS breaking in the chiral cavity can affect the TRS of the ground state of the exciton. In general, a cavity with right-handed polarized photons will decrease the energy of the  $\ell<0$ states while left-handed photons, the time-reversed cavity will decrease the energy of $\ell>0$ states, the time reversed versions of $\ell<0$.

We do not assume any particular dependence of $g$ on $\omega$ in this work because different cavity designs may have differing dependencies of $g$ on $\omega$. For example, in some cavities \cite{JuniCavity}, the dependence of the cavity mode volume on $\omega^{-3}$ results in $\alpha$ independent of $\omega$.

We would like to remark that although previous work has been done on excitons in TMDs in a linear cavity \cite{cavityExciton}, our work is distinct in that our TMDs are separated by an insulating barrier, our cavity is chiral, and we explore energy levels of the long-lived interlayer exciton.

Upon finish we noticed relevant work that shares a similar spirit where these authors focused on the hydrogen atom problem in a chiral cavity\cite{AMDependentShift}. Although, there is similarity, we want to point out some key differences: our work is focused on exciton physics, where the the charge carriers have motion allowed in the x-y plane but constrained in the z-direction by the interlayer separation provided by the insulating barrier. We also show that the cavity does not couple to the center of mass motion except for a  constant shift in energy as seen in equation (\ref{eq: bare Heff}). We suspect that there are links between the two methods and believe that our method is model-independent and can be applied, with some modifications, to a plethora of light-matter interaction Hamiltonians.

In our analysis we point out that the $d\rightarrow 0$ limit cannot be taken since we assume that our TMD layers are separated by an insulating barrier. Work has been done on twisted bilayer graphene in a cavity \cite{tbgCavity} it is perhaps worthwhile to investigate twisted TMDs and their excitons in a chiral cavity. 

It is important to note that the phase boundary provided here is applicable to any interlayer exciton system as long as the Hamiltonian can be approximated as in equation (\ref{eq: H0}) such as double quantum well structure GaAs-AlGaAs \cite{interactionAndCoherence}. This phase boundary serves as an important region of the parameter space at which single-particle states are modified, which is important for many-body studies. The ability to effectively control the angular momentum of the energy spectrum of long-lived interlayer excitons opens the route to tunable single-photon sources \cite{singlePhotonEmitter} and engineered excitonic devices with tailored angular momentum properties.
\section{Acknowledgement}
Z. L. acknowledges helpful discussions with Michael Vogl and Guangyue Ji. This work was supported by the Quantum Science Center (QSC), a National Quantum Information Science Center of the U.S. Department of Energy (DOE). We also acknowledge Grant Number GBMF12976 from the Gordon and Betty Moore Foundation. 

\newpage
\appendix
\bibliography{apssamp}
\newpage
\onecolumngrid 
\section{Unperturbed Hamiltonian:}
We consider an electron and a hole separated by a distance d in a TMD bilayer as in \cite{e-hFluid}, thus our system is effectively 2 dimensional.

Our Hamiltonian in position space is
\begin{align}
    H_0 &= -\frac{\hbar^2 \del_e ^2}{2m_2} -\frac{\hbar^2 \del_e ^2}{2m_h} - \frac{e^2}{4 \pi \epsilon} \frac{1}{\sqrt{(x_e - x_h)^2 + (y_e - y_h)^2 + d^2}}\\
    H_0 &= \frac{p_e^2}{2m_e} + \frac{p_h^2}{2m_h} - \frac{e^2}{4 \pi \epsilon} \frac{1}{\sqrt{(x_e - x_h)^2 + (y_e - y_h)^2 + d^2}} 
\end{align}

Where $\del_e$ is the gradient with respect to the coordinates of the electron and $\del_h$ for the hole. $m_e$ and $m_h$ are the masses of the electron and the hole, respectively.
The goal is to find the spectrum of bound states of our system in the presence of a chiral cavity. We proceed in the following manner:
\begin{itemize}
    \item Introduce a cavity to our system
    \item Transform variables to get an equation in the center of mass frame of the electron-hole pair
    \item Integrate the photonic bath
    \item Convert to cylindrical coordinates
    \item Solve the Schrodinger equation by performing separation of variables
\end{itemize}
We will treat the cavity quantum mechanically so our Hamiltonian is time-independent.
\section{Cavity}
The quantized vector potential for the 0-momentum photon $\mathbf{A}$ is given by \cite{grapheneCavity}
\begin{equation}\label{eq: A linear}
    \mathbf{A} = \sqrt{\frac{\hbar}{\epsilon_C \mathcal{V} 2 \omega _c}}\sum_{\lambda = x,y} \be_\lambda \left(b_\lambda^\dagger + b_\lambda\right)
\end{equation}
With dispersion relation given by \cite{CavitySC}

\begin{equation}
    \omega_q = \omega_c\left(1+\frac{c\abs{\bq}}{\omega_c}\right)^{1/2}
\end{equation}

Where $\bq$ is the in-plane momentum \cite{AQuantization} and $\mathcal{V}$ is the effective mode volume dependent on the cavity design. We take the $\bq = 0$ mode only, hence no dependence on $\bq$ in our cavity vector potential.\\\\
The circular polarization conventions we use are
\begin{align}
\begin{cases}
    \be_L \equiv (1, - i)/\sqrt{2}\\
    \be_R \equiv (1, + i)/\sqrt{2}
\end{cases}
&\Leftrightarrow 
\begin{cases}
    \be_x = (\be_R + \be_L)/\sqrt{2}\\
    \be_y = -i(\be_R - \be_L)/\sqrt{2}
\end{cases}\\ 
\begin{cases}
    b_L = \frac{1}{\sqrt{2}}(b_x - i b_y)\\
    b_R = \frac{1}{\sqrt{2}} (b_x + i b_y)
\end{cases}
&\Leftrightarrow \begin{cases}
    b_x = \frac{1}{\sqrt{2}}(b_L + b_R)\\
    b_y = \frac{i}{\sqrt{2}}(b_L - b_R)
\end{cases}
\end{align}

\section{The $(\hbar k-qA)^2$ term}
Now, we compute $(\bk - q \mathbf{A})^2 = k^2 - 2q \bk \cdot \mathbf{A} + q^2 A^2$
\begin{align}
    A^2 &= \mathbf{A}\cdot \mathbf{A}\\
    &\underbrace{=}_{(\ref{eq: A linear})} \frac{\hbar}{\epsilon_C \mathcal{V}2\omega_c}\left[\left(b_x^\dagger +b_x\right)^2 + \left(b_y^\dagger +b_y\right)^2\right]\\
    &=\frac{\hbar}{\epsilon_C \mathcal{V}2\omega_c} \left[(b_x^\dagger)^2 + b_x^\dagger b_x + \underbrace{b_x b_x^\dagger}_{[b_x, b_x^\dagger] + b_x^\dagger b_x} + b_x^2 + (x\leftrightarrow y)\right]\\
    &= \frac{\hbar}{\epsilon_C \mathcal{V}2\omega_c} \left[(b_x^\dagger)^2 + 2b_x^\dagger b_x + 1 + b_x^2 +(x\leftrightarrow y) \right]\\
    &= \frac{\hbar}{\epsilon_C \mathcal{V}2\omega_c} \left[(b_x^\dagger)^2 + 2b_x^\dagger b_x + b_x^2 +1\right] + \frac{\hbar}{\epsilon_C \mathcal{V}2\omega_c} \left[(b_y^\dagger)^2 + 2b_y^\dagger b_y + b_y^2 + 1\right]   
\end{align}
We will ultimately work in circular polarizations, so we use such coordinates
\begin{align*}
    A^2 = &\frac{\hbar}{\mathcal{V} \epsilon_C 2\omega_c}\left[\left(\frac{b_L^{\dagger}+b_R^{\dagger}}{\sqrt{2}}\right)^2
    + 2\left(\frac{b_L^{\dagger}+b_R^{\dagger}}{\sqrt{2}}\right) \left(\frac{b_L+b_R}{\sqrt{2}}\right) 
    + \left(\frac{b_L+b_R}{\sqrt{2}}\right)^2 
    + 1\right.\\
    &+\left.\left(\left(\frac{i}{\sqrt{2}}\left(b_L-b_R\right)\right)^{\dagger}\right)^2
+2\left(\frac{i}{\sqrt{2}}\left(b_L-b_R\right)\right)^{\dagger}\left(\frac{i}{\sqrt{2}}\left(b_L-b_R\right)\right)+\left(\frac{i}{\sqrt{2}}\left(b_L-b_R\right)\right)^2+1\right]\\
&= \frac{\hbar}{\mathcal{V} \epsilon_C 2\omega_c} \left[\frac{1}{2}\left({b_L^\dagger}^2 + 2b_L^\dagger b_R^\dagger + {b_R^\dagger}^2\right) + \left(b_L^\dagger b_L + b_L^\dagger b_R + b_R^\dagger b_L + b_R^\dagger b_R\right) + \frac{1}{2}\left(b_L^2 + 2b_Lb_R + b_R^2\right) + 2\right]\\
&- \frac{\hbar}{\mathcal{V} \epsilon_C 2\omega_c} \left[\frac{1}{2}\left({b_L^\dagger}^2 - 2b_L^\dagger b_R^\dagger + {b_R^\dagger}^2\right) - \left(b_L^\dagger b_L - b_L^\dagger b_R - b_R^\dagger b_L + b_R^\dagger b_R\right) +\frac{1}{2}\left(b_L^2 - 2 b_L b_R + {b_R}^2\right)\right]\\
&= \frac{\hbar}{\mathcal{V} \epsilon_C 2\omega_c} \left[2 + 2b_L^\dagger b_R^\dagger + 2 b_L b_R + 2 b_L^\dagger b_L + 2 b_R^\dagger b_R\right]\\
&= \frac{\hbar}{\mathcal{V} \epsilon_C 2\omega_c} \left[(1 + 2 b_R^\dagger b_R) + (1 + 2 b_L^\dagger b_L) +2b_L^\dagger b_R^\dagger + 2 b_L b_R \right]
\end{align*}
Using \cite{grapheneCavity},
\begin{equation}
    \mathbf{A} = \sqrt{\frac{\hbar}{\epsilon_C \mathcal{V}2 \omega_c}}\left[\be_R b_L + \be_R b_R^\dagger + \be_L b_R + \be_L b_L^\dagger\right]
\end{equation}
It's not too difficult to show that,
\begin{equation}
    \bk \cdot \mathbf{A} = \sqrt{\frac{\hbar}{\epsilon_C \mathcal{V}2 \omega_c}}\frac{1}{\sqrt{2}}\left[k_x \left( b_L + b_R^\dagger + b_R + b_L^\dagger \right) + i k_y \left(b_L + b_R^\dagger - b_R - b_L^\dagger\right)\right]
\end{equation}
So we have,
\[\begin{aligned}
    (\hbar \bk - q\mathbf{A})^2 = &\left[\hbar^2 k^2\right] - 2q \hbar\sqrt{\frac{\hbar}{2\mathcal{V}\omega_c \epsilon_C}} \frac{1}{\sqrt{2}}\left[b_L (k_x + i k_y) + b_L^\dagger (k_x - i k_y) + b_R (k_x - i k_y) + b_R^\dagger (k_x + ik_y)\right]\\
    & + q^2\frac{\hbar}{2 \mathcal{V}\omega_c \epsilon_C}\left[(1 + 2 b_R^\dagger b_R) + (1 + 2 b_L^\dagger b_L) +2b_L^\dagger b_R^\dagger + 2 b_L b_R \right]
\end{aligned}\]
Let us define
\begin{equation}
    g^2 = \frac{\hbar e^2}{2\mathcal{V}\omega_c \epsilon_C}
\end{equation}
Thus, we have
\[\boxed{\begin{aligned}
    (\hbar \bk - q\mathbf{A})^2 = &\left[\hbar^2 k^2\right] - 2\hbar\mathrm{ sign}(q)g \frac{1}{\sqrt{2}}\left[b_L (k_x + i k_y) + b_L^\dagger (k_x - i k_y) + b_R (k_x - i k_y) + b_R^\dagger (k_x + ik_y)\right]\\
    & + g^2\left[(1 + 2 b_R^\dagger b_R) + (1 + 2 b_L^\dagger b_L) +2b_L^\dagger b_R^\dagger + 2 b_L b_R \right]
\end{aligned}}\]
Our full Hamiltonian is, keeping only Right-handed polarization photons:
\begin{align*}
    H &= H_0(\bp \rightarrow \bp - q\bA) + \hbar \omega_c\sum_{\lambda = R,L}\left( b^\dagger_\lambda b_\lambda + \frac{1}{2}\right)\\
    &= \frac{p_e^2}{2m_e} + \frac{p_h^2}{2m_h} +  \left(\frac{1}{2m_e} + \frac{1}{2m_h}\right) g^2 \left[(1 + 2 b_R^\dagger b_R)  \right]\\
    &+ \hbar \omega_c\left( b^\dagger_R b_R + \frac{1}{2}\right) - \frac{e^2}{4 \pi \epsilon} \frac{1}{\sqrt{(x_e - x_h)^2 + (y_e - y_h)^2 + d^2}}\\
    &+ \frac{g}{\sqrt{2}}\left[ b_R \left(\frac{p^e_x - i p^e_y}{m_e} - \frac{p^h_x - i p^h_y}{m_h}\right)+ b_R^\dagger \left(\frac{p^e_x + ip^e_y}{m_e} - \frac{p^h_x + ip^h_y}{m_h}\right)\right]\\
    &= \frac{1}{2}\hbar \Omega +\frac{p_e^2}{2m_e} + \frac{p_h^2}{2m_h} -  \frac{e^2}{4 \pi \epsilon} \frac{1}{\sqrt{(x_e - x_h)^2 + (y_e - y_h)^2 + d^2}} + \hbar \Omega b_R^\dagger b_R \\
    &+ \frac{g}{\sqrt{2}}\left[ b_R \left(\frac{p^e_x - i p^e_y}{m_e} - \frac{p^h_x - i p^h_y}{m_h}\right)+ b_R^\dagger \left(\frac{p^e_x + ip^e_y}{m_e} - \frac{p^h_x + ip^h_y}{m_h}\right)\right]
\end{align*}
Where:
\begin{align}
    \frac{1}{\mu} &= \frac{1}{m_e} + \frac{1}{m_h}\label{eq: redMass}\\
    \hbar \Omega &= \hbar \omega_c +\frac{g^2}{\mu}
\end{align}
The Hamiltonian can be written as:
\begin{equation}
    H = \zeta + A\; b_R^\dagger b_R + w'b_R^\dagger + w^\dagger b_R\\
\end{equation}
With
\begin{align}
    \zeta &= \frac{1}{2}\hbar \Omega +\frac{p_e^2}{2m_e} + \frac{p_h^2}{2m_h} -  \frac{e^2}{4 \pi \epsilon} \frac{1}{\sqrt{(x_e - x_h)^2 + (y_e - y_h)^2 + d^2}}\\
    A &= \hbar \Omega\\
    w'&= \frac{g}{\sqrt{2}}\left(\frac{p^e_x + ip^e_y}{m_e} - \frac{p^h_x + ip^h_y}{m_h}\right) \label{eq: w'}\\
    w^\dagger &= \frac{g}{\sqrt{2}}\left(\frac{p^e_x - ip^e_y}{m_e} - \frac{p^h_x - ip^h_y}{m_h}\right)\label{eq: w dagger}
\end{align}
\section{Center of Mass Frame}
We perform the following change of variables:
\begin{equation}
    \begin{cases}
    \bs= \mathbf{r}_e - \mathbf{r}_h\\
    \bR = \mu \left(\frac{\br_e}{m_h} + \frac{\br_h}{m_e}\right)\\
    \bP = \bp_e + \bp_h\\
    \bq = \mu\left(\frac{\bp_e}{m_e} - \frac{\bp_h}{m_h}\right)
    \end{cases}
\end{equation}
The inverse of this transformation is
\begin{equation}
    \begin{cases}
    \br_e = \bR + \frac{\mu}{m_e}\bs\\
    \br_h = \bR -\frac{\mu}{m_h}\bs\\
    \bp_e = \frac{\mu}{m_h}\bP+ \bq\\
    \bp_h = \frac{\mu}{m_e}\bP - \bq
    \end{cases}\label{eqs: COM in terms of normal}
\end{equation}
Where $\mu$ is the reduced mass (\ref{eq: redMass}), $\bs$ and $\bq$ are canonically conjugate variables and so are $\bR$ and $\bP$ so that
\begin{equation}
    [R_j, P_k] = [s_j, q_k] = \delta_{jk} i\hbar
\end{equation}
and
\begin{equation}
    [q_j, P_k] = [s_j, R_k] = 0
\end{equation}
Now we have
\begin{align}
    \zeta &= \frac{g}{\mu} + \frac{1}{2} \hbar \omega_c +\frac{\bP^2}{2 (m_e + m_h)}  + \frac{\bq^2}{2\mu}  - \frac{e^2}{4 \pi \epsilon}\frac{1}{\sqrt{\bs^2 + d^2}}\\
    A&= \hbar \Omega\\
    w' &= \frac{g}{\sqrt{2}\mu}\left(q_x + i q_y\right)\\
    w^\dagger &= \frac{g}{\sqrt{2}\mu}\left(q_x - i q_y\right)
\end{align}
\section{Path Integral}
If a single-body Hamiltonian can be separated into $H = T(p) + V(x)$, then the partition function can be written as \cite{A&S}:
\begin{equation}
    Z \simeq\int dQ\int \prod_{\substack{n=1 \\ q_N=Q=q_0}}^{N-1} d q_n \prod_{n=1}^N \frac{d p_n}{2 \pi \hbar}
    \exp{\frac{-\Delta \tau}{\hbar} \sum_{n=0}^{N-1}\left(V\left(q_n\right)+T\left(p_{n+1}\right)-ip_{n+1} \frac{q_{n+1}-q_n}{\Delta \tau}\right)}
\end{equation}
Where $\Delta \tau = \beta\hbar/N$. Taking the limit as $N\rightarrow\infty$:
\begin{equation}
    Z \simeq\int \mathcal{D}q \mathcal{D}p
    \exp\left\{\frac{-1}{\hbar} \int_0^{\beta\hbar} d\tau\left[H(p,q)-ip \partial_\tau q\right]\right\}
\end{equation}
where 
\begin{align}
    \mathcal{D}q = \lim_{N\rightarrow\infty}\prod_{\substack{n=1 \\ q_N=Q=q_0}}^{N-1} d q_n\\
    \mathcal{D}p = \lim_{N\rightarrow\infty} \left(\frac{1}{2\pi\hbar}\right)^N\prod_{n-1}^Ndp_n
\end{align}
 In our case we must also address the many-body photonic field. The photonic field, the electron and the hole are distinct particles, so we may write the full partition function as:
 \begin{equation}
     Z \simeq\int \mathcal{D}q \mathcal{D}p \int \mathcal{D}\phi
    \exp{\frac{-1}{\hbar} \int_0^{\beta\hbar} d\tau\left[\frac{\hbar}{2}\phi^*(\overrightarrow{\partial_\tau} - \overleftarrow{\partial_\tau})\phi +H(p,q,\phi)-ip \partial_\tau q\right]}
 \end{equation}
with
\begin{equation}
    \mathcal{D}\phi = \lim_{N\rightarrow\infty}\prod_{\substack{n=1 \\ \phi_0 = \phi_N}}^Nd\phi_n.
\end{equation}
Here, the notation $A\overrightarrow{\partial_\tau}B \equiv A(\partial_\tau B)$, and $A\overleftarrow{\partial_\tau}B \equiv (\partial_\tau A)B$. This can be obtained from the more commonly used form 
\begin{equation}
         Z \simeq\int \mathcal{D}q \mathcal{D}p \int \mathcal{D}\phi
    \exp{\frac{-1}{\hbar} \int_0^{\beta\hbar} d\tau\left[\hbar\phi^* \partial_\tau\phi +H(p,q,\phi)-ip \partial_\tau q\right]}
\end{equation}
by subtracting the total derivative $\frac{\hbar}{2}\partial_\tau(\phi^* \phi)$ from the integrand. The integrand is the action and adding total time derivatives to the action results in no change to the equations of motion and thus no change to observables. This rewriting is needed in order to obtain a hermitian effective Hamiltonian as will be remarked later.

We can focus on the photonic part of this integration,
\begin{align}
    Z &= \int \mathcal{D}x \;\mathcal{D}(\phi^*_R, \phi_R)e^{-\frac{1}{\hbar} \int d\tau \frac{\hbar}{2}\phi^*_R(\overrightarrow{\partial_\tau} - \overleftarrow{\partial_\tau})\phi_R + \zeta + A\; \phiRstar \phiR + w'\phiRstar + w^\dagger \phiR -ip\partial_\tau q}\\
    &= \int \mathcal{D}x\;\mathcal{D}(\phi^*_R, \phi_R)e^{-\frac{1}{\hbar} \int d\tau \phi^*_R( A + \frac{\hbar}{2}\overrightarrow{\partial_\tau} - \frac{\hbar}{2}\overleftarrow{\partial_\tau})\phi_R  + w'\phiRstar + w^\dagger \phiR + \zeta-ip\partial_\tau q}\\
    &= \int \mathcal{D}x\;e^{-\frac{1}{\hbar}\int d\tau\zeta }\cdot \frac{e^{+\int d\tau \frac{w^\dagger}{\hbar}(\Omega + \frac{1}{2}\overrightarrow{\partial_\tau} - \frac{1}{2}\overleftarrow{\partial_\tau})^{-1}\frac{w'}{\hbar}-ip\partial_\tau q}}{\det(\Omega + \frac{1}{2}\overrightarrow{\partial_\tau} - \frac{1}{2}\overleftarrow{\partial_\tau})}
\end{align}
The determinant will be just a multiplicative constant, which does not affect the partition function. To 1st order we have
\begin{equation}
    \frac{1}{\Omega}\frac{1}{(1 + \frac{1}{2\Omega}\overrightarrow{\partial_\tau} - \frac{1}{2\Omega}\overleftarrow{\partial_\tau})} \approx \frac{1}{\Omega}\left(1- \frac{1}{2\Omega}\overrightarrow{\partial_\tau} + \frac{1}{2\Omega}\overleftarrow{\partial_\tau}\right)
\end{equation}
We now combine both exponentials and obtain
\begin{equation}
    Z \sim e^{-\frac{1}{\hbar}\int d\tau \; \zeta - w^\dagger\frac{1}{\hbar \Omega}\left(1- \frac{1}{2\Omega}\overrightarrow{\partial_\tau} + \frac{1}{2\Omega}\overleftarrow{\partial_\tau}\right) w'-ip\partial_\tau q}
\end{equation}
The effective Hamiltonian is then
\begin{equation}
    H_{\mathrm{eff}} = \zeta -\frac{1}{\hbar\Omega} w^\dagger w' + \frac{1}{2\hbar \Omega^2}w^\dagger \left(\overrightarrow{\partial_\tau} - \overleftarrow{\partial_\tau}\right)w'
\end{equation}

We need to calculate terms of the form $\partial_\tau p_j^\alpha$. These can be evaluated by recalling that in going from propagator to partition function, the substitution made is $t = -i\tau$, so that the Heisenberg equation for operators
\begin{equation}
    \frac{d}{idt}p_j^\alpha = \frac{[H, p_j^\alpha]}{\hbar} \Leftrightarrow \frac{d}{d\tau}p_j^\alpha =\frac{[H, p_j^\alpha]}{\hbar}
\end{equation}
Our Hamiltonian is now
\begin{equation}
    H_{\mathrm{eff}} = \frac{g}{\mu} + \frac{1}{2} \hbar \omega_c +\frac{\bP^2}{2 (m_e + m_h)}  + \frac{\bq^2}{2\mu}  - \frac{e^2}{4 \pi \epsilon}\frac{1}{\sqrt{\bs^2 + d^2}} -\frac{1}{\hbar \Omega}\frac{g^2}{2\mu^2}\bq^2 + \frac{1}{2\hbar \Omega^2}\left[w^\dagger (\partial_\tau w') - (\partial_\tau w^\dagger)w' \right]
\end{equation}
Let's now evaluate the imaginary time derivatives:
\subsection{Imaginary Time derivatives}
We seek to evaluate
\begin{equation}
    \tilde{H} \equiv \frac{1}{2\hbar \Omega^2}\left[w^\dagger (\partial_\tau w') - (\partial_\tau w^\dagger)w' \right]
\end{equation}
where, as discussed,
\begin{equation}
    \frac{d}{d\tau}p_j^\alpha =\frac{[H, p_j^\alpha]}{\hbar}
\end{equation}
The only term that does not commute with momenta in our Hamiltonian is the coulomb interaction. Thus, we seek to evaluate:
\begin{equation}
    \frac{-e^2}{4\pi\epsilon}\left[\frac{1}{\sqrt{\bs^2 + d^2}}, q_j\right]
\end{equation}
This is not too difficult, the result is:
\begin{align}
    \frac{-e^2}{4\pi\epsilon}\left[\frac{1}{\sqrt{\bs^2 + d^2}}, q_j\right] &= \frac{-i\hbar e^2}{4\pi\epsilon}\frac{d}{ds_j}\frac{1}{\sqrt{\bs^2 + d^2}}\\
    &= \frac{i\hbar e^2}{4\pi\epsilon}\frac{s_j}{\sqrt{\bs^2 + d^2}}
\end{align}
Thus
\begin{align}
    \tilde{H} &=\frac{i}{2\hbar \Omega^2}\left[\frac{g}{\sqrt{2}\mu}\left(q_x - i q_y\right) \frac{g}{\sqrt{2}\mu}\frac{ e^2}{4\pi\epsilon}\left(\frac{s_x+is_y}{\sqrt{\bs^2 + d^2}^3}\right) - 
     \frac{g}{\sqrt{2}\mu}\frac{ e^2}{4\pi\epsilon}\left(\frac{s_x-is_y}{\sqrt{\bs^2 + d^2}^3}\right)\frac{g}{\sqrt{2}\mu}\left(q_x + i q_y\right)\right]\\
     &= \frac{i}{\hbar}\frac{e^2}{4\pi\epsilon}\left(\frac{g}{2\mu\Omega}\right)^2\left[(q_x-iq_y) \left(\frac{s_x+is_y}{\sqrt{\bs^2 + d^2}^3}\right) - \left(\frac{s_x-is_y}{\sqrt{\bs^2 + d^2}^3}\right) (q_x+iq_y)\right]\\
\end{align}
Let's focus on the term in brackets.
\begin{align*}
    \left[(q_x-iq_y) \left(\frac{s_x+is_y}{\sqrt{\bs^2 + d^2}^3}\right) - \left(\frac{s_x-is_y}{\sqrt{\bs^2 + d^2}^3}\right) (q_x+iq_y)\right] &= \left[q_x, \frac{s_x}{(\bs^2 + d^2)^{3/2}}\right] -i^2\left[q_y, \frac{s_y}{(\bs^2 + d^2)^{3/2}}\right]\\
    &+i \left(q_x \frac{s_y}{(\bs^2 + d^2)^{3/2}} + \frac{s_y}{(\bs^2 + d^2)^{3/2}} q_x\right)\\
    &-i\left( q_y \frac{s_x}{(\bs^2 + d^2)^{3/2}} + \frac{s_x}{(\bs^2 + d^2)^{3/2}}q_y\right)\\
    &= \left[q_x, \frac{s_x}{(\bs^2 + d^2)^{3/2}}\right] +\left[q_y, \frac{s_y}{(\bs^2 + d^2)^{3/2}}\right]\\
    &- 2i\frac{1}{(\bs^2 + d^2)^{3/2}} (s_x q_y - s_y q_x)\\
    &= \left[q_x, \frac{s_x}{(\bs^2 + d^2)^{3/2}}\right] +\left[q_y, \frac{s_y}{(\bs^2 + d^2)^{3/2}}\right]\\
    &-\frac{2i}{(\bs^2 + d^2)^{3/2}}\mathbf{L}\cdot \mathbf{\hat z}\\
    &= i\hbar \frac{\bs^2-2d^2}{(\bs^2 + d^2)^{5/2}}-\frac{2i}{(\bs^2 + d^2)^{3/2}}\mathbf{L}\cdot \mathbf{\hat z}
\end{align*}

Where we have \begin{equation}
    \mathbf{L} = \bs\times \bq
\end{equation}
Which corresponds to the relative angular momentum. The minus sign on the LHS of the above equation arises from the rewriting of the action in the path integral. Without this rewriting we would only have the first term, which is not hermitian.\\
Thus, we obtain
\begin{align}
    \tilde{H} = \frac{1}{\hbar}\frac{e^2}{4\pi\epsilon}\left(\frac{g}{2\mu\Omega}\right)^2\left[\hbar \frac{2d^2 - \bs^2}{(\bs^2 + d^2)^{5/2}}+\frac{2}{(\bs^2 + d^2)^{3/2}}\mathbf{L}\cdot \mathbf{\hat z}\right]
\end{align}
So our effective Hamiltonian is
\begin{align*}
    H_{\mathrm{eff}} &=\frac{g}{\mu} + \frac{1}{2} \hbar \omega_c +\frac{\bP^2}{2 (m_e + m_h)}  + \frac{\bq^2}{2\mu}  - \frac{e^2}{4 \pi \epsilon}\frac{1}{\sqrt{\bs^2 + d^2}} -\frac{1}{\hbar \Omega}\frac{g^2}{2\mu^2}\bq^2\\
    &+ \frac{1}{\hbar}\frac{e^2}{4\pi\epsilon}\left(\frac{g}{2\mu\Omega}\right)^2\left[\hbar \frac{2d^2 - \bs^2}{(\bs^2 + d^2)^{5/2}}+\frac{2}{(\bs^2 + d^2)^{3/2}}\mathbf{L}\cdot \mathbf{\hat z}\right]
\end{align*}
Define:
\begin{equation}
    \Sigma = \mu +\frac{g^2}{\hbar\omega_c} = \frac{\mu\Omega}{\omega_c}
\end{equation}
Then our Hamiltonian can be written as 
\begin{align*}
    H_{\mathrm{eff}} &=\frac{g}{\mu} + \frac{1}{2} \hbar \omega_c +\frac{\bP^2}{2 (m_e + m_h)}  + \frac{\bq^2}{2\Sigma}  - \frac{e^2}{4 \pi \epsilon}\frac{1}{\sqrt{\bs^2 + d^2}}\\
    &- \frac{e^2}{4\pi\epsilon}\left(\frac{g}{2\Sigma \omega_c}\right)^2\left[ \frac{\bs^2-2d^2}{(\bs^2 + d^2)^{5/2}}-\frac{2}{(\bs^2 + d^2)^{3/2}}\frac{\mathbf{L}}{\hbar}\cdot \mathbf{\hat z}\right]
\end{align*}
Let's show that the last term is Hermitian. It is proportional to
\begin{equation}
    \frac{1}{(\bs^2 + d^2)^{3/2}}\left(s_xq_y - s_yq_x\right)
\end{equation}
Note that
\begin{align}
    \left[\frac{1}{(\bs^2 + d^2)^{3/2}}\left(s_xq_y - s_yq_x\right)\right]^\dagger &= \left(q_ys_x - q_xs_y\right)\frac{1}{(\bs^2 + d^2)^{3/2}}\\
    &= \left(s_xq_y - s_yq_x\right)\frac{1}{(\bs^2 + d^2)^{3/2}}\\
    &= s_x\left(\frac{1}{(\bs^2 + d^2)^{3/2}}q_y + \left[q_y,\frac{1}{(\bs^2 + d^2)^{3/2}}\right]\right)\nonumber\\
    &- s_y\left(\frac{1}{(\bs^2 + d^2)^{3/2}}q_x + \left[q_x,\frac{1}{(\bs^2 + d^2)^{3/2}}\right]\right)\\
    &= \frac{1}{(\bs^2 + d^2)^{3/2}}s_xq_y + s_x\left[q_y,\frac{1}{(\bs^2 + d^2)^{3/2}}\right]\nonumber\\
    &- \frac{1}{(\bs^2 + d^2)^{3/2}}s_yq_x - s_y\left[q_x,\frac{1}{(\bs^2 + d^2)^{3/2}}\right]\\
    &= \frac{1}{(\bs^2 + d^2)^{3/2}}\left(s_xq_y - s_yq_x\right)\nonumber\\
    &+ s_y \left[q_x,\frac{1}{(\bs^2 + d^2)^{3/2}}\right] - s_y\left[q_x,\frac{1}{(\bs^2 + d^2)^{3/2}}\right]\\
    &= \frac{1}{(\bs^2 + d^2)^{3/2}}\left(s_xq_y - s_yq_x\right)
\end{align}
Where we in the last line we have used that difference in commutators can be written as
\begin{equation}
    -i\hbar \sum_{i,j}\epsilon_{ij}s_i\partial_j\frac{1}{(\bs^2 + d^2)^{3/2}}
\end{equation}
and noting that
\begin{equation}
    s_i\partial_j\frac{1}{(\bs^2 + d^2)^{3/2}} = s_j\partial_i\frac{1}{(\bs^2 + d^2)^{3/2}}
\end{equation}

Thus, we obtain a hermitian Hamiltonian. 
\section{Separation of  variables}
Define:
\begin{equation}
    M = m_e + m_h
\end{equation}
Let's assume that our wavefunction can be separated as:
\begin{equation}
    \Psi(s,R) = \phi_1(R)\phi_2(s)
\end{equation}
Then,
\begin{align*}
    \left(E-\frac{g}{\mu} + \frac{1}{2}\hbar \omega_c\right)\Psi(\bs,\bR) &= H \Psi(\bs,\bR)\\
    &= H \phi_1(\bR)\phi_2(\bs)\\
    &= -\frac{\hbar^2}{2M}\phi_2(\bs)\del^2_\bR\phi_1(\bR) -\frac{\hbar^2}{2\Sigma}\phi_1(\bR)\del^2_\bs\phi_2(\bs)\\
    & - \phi_1(\bR)\frac{e^2}{4 \pi \epsilon}\frac{1}{\sqrt{s^2 + d^2}}\phi_2(\bs)
    + \phi_1(\bR)\frac{1}{\hbar}\frac{e^2}{4\pi\epsilon}\left(\frac{g}{2\Sigma \omega_c}\right)^2\left[\hbar \frac{2d^2 - \bs^2}{(\bs^2 + d^2)^{5/2}}+\frac{2}{(\bs^2 + d^2)^{3/2}}\mathbf{L}\cdot \mathbf{\hat z}\right]\phi_2(\bs)
\end{align*}
As is customary, we divide both sides by $\Psi(s,R)$ and rearrange. By redefining $E$ as $E-\frac{g}{\mu} + \frac{1}{2}\hbar \omega_c$ we obtain
\begin{align}
    -\frac{\hbar^2}{2 \Sigma}\frac{\del^2_s \phi_2(\bs)}{\phi_2(\bs)}
    -\frac{e^2}{4 \pi \epsilon}\frac{1}{\sqrt{\bs^2 + d^2}}\left\{1- \frac{1}{\hbar}\frac{1}{\bs^2 + d^2}\left(\frac{g}{2\Sigma \omega_c}\right)^2\left[\hbar \frac{2d^2 - \bs^2}{\bs^2 + d^2}+\frac{2}{\phi_2(\bs)}\mathbf{L}\cdot \mathbf{\hat z}\;\phi_2(\bs)\right]\right\} &= E + \underbrace{\frac{\hbar^2}{2M}\frac{\del^2 \phi_1(\bR)}{\phi_1(\bR)}}_{-E^R}\\
    &=:E^s
\end{align}
We denote $-\frac{\hbar^2}{2M}\frac{\del^2 \phi_1(R)}{\phi_1(R)}$ as $E^R$, it corresponds to the center of mass kinetic energy. This equality can only hold if both sides are constants because the LHS is a function of only s and the RHS is a function of only R. 
Our Schrodinger equation for $s$, the separation is
\begin{equation}
    \frac{\hbar^2}{2 \Sigma}\del^2_s \phi_2(\bs) -\frac{e^2}{4 \pi \epsilon}\frac{1}{\sqrt{\bs^2 + d^2}}\left\{1- \frac{1}{\hbar}\frac{1}{\bs^2 + d^2}\left(\frac{g}{2\Sigma \omega_c}\right)^2\left[\hbar \frac{2d^2 - \bs^2}{\bs^2 + d^2}+2\mathbf{L}\cdot \mathbf{\hat z}\right]\right\}\phi_2(\bs) = E^s\phi_2(\bs)
\end{equation}
Where $E^s$ is the bound-state contribution to the total energy.
\section{Cylindrical coordinates}
In cylindrical coordinates ($s$, $\theta$), we obtain:
\begin{align}
    E^s \phi_2(\bs) = &-\frac{\hbar^2}{2\Sigma}\left[\frac{1}{s}\frac{\partial}{\partial s}\left(s\frac{\partial}{\partial s}\right)+\frac{1}{s^2}\frac{\partial^2}{\partial \theta^2}\right]\phi_2(s)\nonumber\\
    &- \frac{e^2}{4 \pi \epsilon}\frac{1}{\sqrt{s^2 + d^2}}\left\{1- \frac{1}{\hbar}\frac{1}{s^2 + d^2}\left(\frac{g}{2\Sigma \omega_c}\right)^2\left[\hbar \frac{2d^2 - s^2}{s^2 + d^2}-2i\hbar \frac{\partial}{\partial \theta}\right]\right\}\phi_2(\bs)\label{eq: cyl coord SE}
\end{align}
We assume that $\phi_2(\bs)$ is of the form,
\begin{equation}
    \phi_2(\bs) = S(s)e^{i\ell \theta}, \;\;\ell\in \{0,\pm 1, \pm 2, ... \}
\end{equation}
We obtain, by multiplying both sides by $e^{-i\ell \theta}$ on the left of Eq. \ref{eq: cyl coord SE},

\begin{align}
    E^s S = &-\frac{\hbar^2}{2\Sigma}\left[\frac{1}{s}\frac{\partial}{\partial s}\left(s\frac{\partial}{\partial s}\right)S-\frac{\ell^2}{s^2}S\right]\nonumber\\
    &- \frac{e^2}{4 \pi \epsilon}\frac{S}{\sqrt{s^2 + d^2}}\left\{1- \frac{1}{\hbar}\frac{1}{s^2 + d^2}\left(\frac{g}{2\Sigma \omega_c}\right)^2\left[\hbar \frac{2d^2 - s^2}{s^2 + d^2}+2\ell\hbar \right]\right\}
\end{align}

Rearranging, we obtain,
\begin{equation}
    -S''(s) - \frac{1}{s}S'(s) - \frac{2\Sigma}{\hbar^2}\left\{E^s + \frac{e^2}{4\pi \epsilon}\frac{1}{\sqrt{s^2 + d^2}}\left[1- \frac{1}{s^2 + d^2}\left(\frac{g}{2\Sigma \omega_c}\right)^2\left(\frac{2d^2 - s^2}{s^2 + d^2}+2\ell\right)\right] - \frac{\hbar^2}{2 \Sigma}\frac{\ell^2}{s^2}\right\}S(s)=0
\end{equation}
Define:
\begin{align}
1Ry^* &= \frac{\Sigma e^4}{2(4\pi \epsilon \hbar)^2}\\
1a^*_B &= \frac{4\pi \epsilon\hbar^2}{\Sigma e^2}\\
\end{align}
The conversion from $Ry^*$ and $a_B^*$ to eV and nm is nontrivial in the following sense: one can modify $\Sigma$ in many ways, by changing the dielectrics, the cavity frequency or the mode volume. Unless parameters are changed in a way such that $g^2/\hbar\omega_c$ is constant, the definition of $Ry^*$ and $a_B^*$ will change with $\alpha$ due to their $\Sigma$-dependence.

Define
\begin{align}
s&=a^*_B x\\
d &= a_B^*D
\end{align}
Then, $S(s) = S(a^*_B x)$ and $\frac{d}{ds} = \frac{d}{a^*_Bdx}$. It is useful to know the following equivalencies:
\begin{align}
    Ry^*(a_B^*)^2 &= \frac{\hbar^2}{2\Sigma}\\
    Ry^*a^*_B &= \frac{1}{2}\frac{e^2}{4\pi \epsilon}
\end{align}
Thus,
\begin{align}
    &-\frac{\partial_x^2S(a_B^*x)}{(a_B^*)^2} - \frac{1}{a_B^*x}\frac{\partial_x S(a^*_Bx)}{a_B^*} - \frac{1}{Ry^* (a_B^*)^2} \left\{E^s  -\frac{Ry^* (a_B^*)^2}{(a_B^*)^2} \frac{\ell^2}{x^2}\right. \nonumber\\
    &\left.  + 2a_B^* Ry\frac{1}{a_B^* \sqrt{x^2 + D^2}}\left[1- \frac{1}{x^2 + D^2}\left(\frac{g}{2\Sigma \omega_c a_B^*}\right)^2\left(\frac{2D^2 - x^2}{x^2 + D^2}+2\ell\right)\right]\right\}S(a^*_B x) = 0\\
    \Rightarrow &\frac{1}{(a_B^*)^2}\left\{\partial_x^2 S(a_B^* x) + \frac{\partial_xS(a_B^* x)}{x}\right.\nonumber\\
    &\left.+ \left[\frac{E^s}{Ry}+\frac{2}{\sqrt{x^2 + D^2}}\left[1- \frac{1}{x^2 + D^2}\left(\frac{g}{2\Sigma \omega_c a_B^*}\right)^2\left(\frac{2D^2 - x^2}{x^2 + D^2}+2\ell\right)\right]-\frac{\ell^2}{x^2}\right]S(a_B^* x)\right\}= 0
\end{align}
Let $X(x) = S(a_B^*x)$, so that $X$ is simply $S$ in rescaled units, if energy is in units of $Ry$ (which we denote as $\xi^s$), our differential equation takes the form:
\begin{equation}\label{eq: the DFQ}
    X''(x) + \frac{X'(x)}{x} + \left[\xi^s + \frac{2}{\sqrt{x^2 + D^2}} - \frac{\ell^2}{x^2} -\frac{1}{2}\left(\frac{g}{\Sigma \omega_c a}\right)^2 \frac{1}{(x^2 + D^2)^{3/2}} \left(2\ell +  \frac{2D^2- x^2}{x^2+D^2}\right)\right]X(x) = 0
\end{equation}
Define
\begin{equation}
    \alpha = \frac{1}{2}\left(\frac{g}{\Sigma \omega_c a}\right)^2
\end{equation}
which is a dimensionless parameter, it's value determines the effects of the cavity on the exciton spectrum.
To solve \ref{eq: the DFQ} numerically we let $\Xi(x) = X'(x)$ so that 
\begin{equation}
    \frac{d}{dx}\begin{bmatrix}
    X\\\Xi
    \end{bmatrix}=\begin{bmatrix}
        0&1\\
        \frac{\ell^2}{x^2} +\frac{1}{2}\left(\frac{g}{\Sigma \omega_c a}\right)^2 \frac{1}{(x^2 + D^2)^{3/2}} \left(2\ell +  \frac{2D^2- x^2}{x^2+D^2}\right) - \frac{2}{\sqrt{x^2 + d^2}} - \xi^s & -\frac{1}{x}
    \end{bmatrix}
    \begin{bmatrix}
        X\\
        \Xi
    \end{bmatrix}
\end{equation}

We solve for the energy levels and eigenstates using a Runge-Kutta integrator
to 4th order (RK4). We integrate from a large electron-hole separation, $x$,  of $35$ down to $10^{-14}$.


\end{document}